\documentclass[a4paper]{jpconf}
\usepackage{graphicx}
\usepackage{amsbsy}
\usepackage{amsfonts}
\usepackage{amssymb}
\usepackage{iopams}
\usepackage{cite}
\begin{document}
\title{Electron impact ionization of stored \\ highly charged ions}

\author{Michael Hahn}

\address{Columbia Astrophysics Laboratory, Columbia University, 550 West 120th Street, New York, NY 10027, USA}
\ead{mhahn@astro.columbia.edu}

\begin{abstract}

Accurate cross section data for electron impact ionization (EII) are needed in order to interpret the spectra of collisionally ionized plasmas both in astrophysics and in the laboratory. Models and spectroscopic diagnostics of such plasmas rely on accurate ionization balance calculations, which depend, in turn, on the underlying rates for EII and electron-ion recombination. EII measurements have been carried out using the TSR storage ring located at the Max-Planck-Institut f\"ur Kernphysik in Heidelberg, Germany. Storage ring measurements are largely free of metastable contamination, resulting in unambiguous EII data, unlike what is encountered with other experimental geometries. As it is impractical to perform experiments for every ion, theory must provide the bulk of the necessary EII data. In order to guide theory, TSR experiments have focused on providing at least one measurement for every isoelectronic sequence. EII data have been measured for ions from 13 isoelectronic sequences: Li-like silicon and chlorine, Be-like sulfur, B-like magnesium, and F-like through K-like iron. These experimental results provide an important benchmark for EII theory. 

\end{abstract}

\section{Introduction}

	Accurate electron impact ionization (EII) cross section data are needed to interpret the spectra of collisionally ionized plasmas. Such plasmas are formed in astrophysical objects such as stars, supernova remnants, galaxies, and galaxy clusters; as well as in laboratory plasmas, such as discharges and tokamaks. Models and spectroscopic diagnostics for these objects rely on accurate ionization balance calculations. The charge state distribution (CSD) depends, in turn, on the ionization and recombination rate coefficients, which are derived from the cross sections. To see this consider the case of collisional ionization equilibrium (CIE), which is valid when the electron density is low, radiation can be ignored, and there is no dust. In CIE, provided there are no multi-electron processes, the ionization and recombination rates are the same so that 
\begin{equation}	
n_{\mathrm{e}} n_{q} \alpha_{\mathrm{I}}^{q} = n_{\mathrm{e}}n_{q+1} \alpha_{\mathrm{R}}^{q+1}, 
\label{eq:equalrates}
\end{equation}
where $n_{\mathrm{e}}$ is the electron density, $n_{q}$ is the density of a given element with charge $q$, $\alpha_{\mathrm{I}}^{q}$ is the rate coefficient for ionization from $q$ to $q+1$ and $\alpha_{\mathrm{R}}^{q+1}$ is the rate coefficient for recombination from $q+1$ to $q$. This expression can be rewritten as 
\begin{equation}
\frac{n_{q+1}}{n_{q}} = \frac{\alpha_{\mathrm{I}}^{q}}{\alpha_{\mathrm{R}}^{q+1}}, 
\label{eq:rateratio}
\end{equation}
which shows that the charge balance depends directly on the ionization and recombination rate coefficients. 
	
	For astrophysical applications, data are needed for all of the elements from hydrogen through zinc \cite{Dere:AA:2007}. For laboratory plasmas, data are also needed for additional elements, such as tungsten \cite{Peacock:CanJPhys:2008,Putterich:NucFus:2010, Schippers:PRA:2011}. Because of the enormous amount of data required, most EII cross sections are calculated theoretically. However, theoretical calculations require approximations to make the problem tractable. Accurate experimental data are essential in order to benchmark theory. 
	
	One of the major limitations for experimental EII studies, is that ion beams usually contain an unknown number of ions in metastable states. Since the ionization cross section for the metastables generally differs from that for ground state ions, the resulting measurements are ambiguous. 
	
	EII measurements using an ion storage ring can overcome this problem. With this approach, the ions are stored for enough time so that metastable levels radiatively decay to the ground state before data are collected. In this way measurements are performed on essentially pure ground state ion beams resulting in unambiguous cross section data. 
	
\section{Experimental method}

	The EII measurements reviewed here were performed using the TSR heavy ion storage ring at the Max-Planck-Insitut f\"ur Kernphysik in Heidelberg, Germany \cite{Habs:NIMB:1989,Grieser:EPJ:2012}. The procedures for the experiment and analysis have been described in detail previously \cite{Linkemann:fe15, Linkemann:fe15a, Linkemann:thesis, Kenntner:cl14, Kenntner:thesis,Hahn:mg7,Hahn:fe11, Hahn:fe12, Hahn:s12, Hahn:fe9, Hahn:fe13, Hahn:fe7, Bernhardt:fe14}. Briefly, a beam of ions of a given mass and charge is injected into TSR. The ion beam is merged to be co-linear with a cold electron beam. Initially, the electron beam velocity is set equal to that of the ions so that elastic electron-ion collisions cool and reduce the energy spread of the ion beam \cite{Poth:PhysRep:1990}. During this initial cooling stage, the metastable levels in the ion beam also decay. 
	
	After cooling, the relative energy of the electrons is varied in steps in order to measure electron-ion collisions. Products of ionizing or recombining collisions are separated from the parent ion beam by a dipole magnet downstream of the interaction region. Detectors are positioned to intercept and count the products. In between each measurement step the background rate is measured at a fixed reference energy. For ionization studies, the reference point is usually chosen to be below the ionization threshold so that the background rate is due only to electron stripping off the residual gas in the vacuum system. The ion and electron current are also measured at each step.
	
	The ionization cross section $\sigma_{\mathrm{I}}$ is determined from these measured quantities. A simplified expression for $\sigma_{\mathrm{I}}$ is
\begin{equation}
\sigma_{\mathrm{I}}(E) = \frac{R(E) - R(E_{\mathrm{r}})}{ v_{\mathrm{rel}}[1-\beta_{i}\beta_{e}(E)] n_{\mathrm{e}}(E) N_{\mathrm{i}} L}.
\label{eq:rate}
\end{equation}	
Here $R(E)$ and $R(E_{\mathrm{r}})$ are the rate at the measurement energy $E$ and the background rate at the reference energy $E_{\mathrm{r}}$, respectively; $v_{\mathrm{rel}}$ is the relative velocity between the electrons and ions; $[1-\beta_{\mathrm{i}}\beta_{\mathrm{e}}]$ is a relativistic correction where $\beta_{\mathrm{i}}$ and $\beta_{\mathrm{e}}$ are the ion and electron velocities normalized by the speed of light; $n_{\mathrm{e}}(E)$ is the electron density, which is derived from the measured electron current; $N_{\mathrm{i}}$ is the number of ions per unit length, which is derived from the ion current; and $L = 1.5$~m is the length of the interaction region. This expression neglects some additional corrections, such as for the background rate when $E_{\mathrm{r}}$ is above the ionization threshold, and for the merging and de-merging sections of the interaction region where the beams are not co-aligned. These and other issues are described in Refs.~\cite{Hahn:mg7,Hahn:fe11} and references therein.

	The typical $1\sigma$ systematic uncertainty of the measured cross section is about 10 -- 15\%. The main source of this uncertainty is the determination of the ion current, which must be measured without destroying the circulating ion beam. This is done with a beam profile monitor (BPM) \cite{Hochadel:NIMA:1994}. The dominant uncertainty in the cross section is fixed by the estimated accuracy of the BPM calibration. Systematic uncertainties from other sources are typically only a few percent, and the uncertainty from counting statistics is usually also very small. 
	
\section{Results}\label{sec:res}

	Here we summarize the major results of the TSR EII studies. These results for each particular measurement are described in more detail in the references cited. Throughout this paper, ions are labelled according to their initial charge state before ionization. 
	
	We compare the TSR measurements to the recommended data of Arnaud \& Rothenflug \cite{Arnaud:AAS:1985}, Arnaud \& Raymond \cite{Arnaud:ApJ:1992}, and Dere \cite{Dere:AA:2007}, because these reviews are the ones most often used for astrophysical studies. The Arnaud \& Rothenflug and Arnaud \& Raymond compilations are largely based on theoretical distorted wave calculations (e.g.,~\cite{Younger:JQSRT:1982, Younger:JQSRT:1983, Pindzola:NucFus:1987}) and some experiments (e.g.,~\cite{Gregory:PRA:1986, Gregory:PRA:1987}). The widely used CIE calculations of Arnaud \& Raymond \cite{Arnaud:ApJ:1992} and Mazzotta et al. \cite{Mazzotta:AAS:1998} are based on those recommended data. The Dere \cite{Dere:AA:2007} cross sections discussed here were calculated using the flexible atomic code (FAC) \cite{Gu:Can:2008}. These EII data have been the basis for the recent CIE calculations by Bryans et al. \cite{Bryans:ApJ:2009}. 

\subsection{Li-like ions $(2s)$}\label{subsec:li}

	EII for the Li-like ions Si$^{11+}$ and Cl$^{14+}$ has been measured at TSR \cite{Kenntner:cl14,Kenntner:thesis}. However, Li-like ions have no low lying metastable levels. Hence crossed beams techniques are also reliable for these ions and there are numerous measurements from such experiments \cite{Muller:PRL:1988a, Hofmann:ZPhysD:1990,Muller:PRA:2000,Teng:PRA:2000}. The TSR measurements found that the predicted direct ionization cross sections were about 20\% larger than the measurement, but otherwise found good agreement with theory. 

\subsection{Be-like ions $(2s^{2})$}\label{subsec:be}

	It is difficult to produce a beam of ground state Be-like ions. The ground level of such ions is $2s^2\,^{1}S_{0}$, but essentially all production methods generate beams with more than half the ions in the $2s\,2p^{\;}\,^{3}P_{0,1,2}$ levels. The $^{3}P_{0}$ level is particularly difficult to remove as it cannot decay through a single photon transition and so has a lifetime of days to centuries, depending on the charge state and nuclear spin \cite{Schmieder:PRA:1973}. 
	
	EII was studied for Be-like S$^{12+}$ \cite{Hahn:s12}. In order to remove the $^{3}P_{0}$ metastable levels the $^{33}$S isotope was used. This isotope has a nuclear spin and the resulting hyperfine interaction reduces the metastable lifetime to $10.4 \pm 0.5$~s \cite{Schippers:PRA:2012}. Thus, by first storing the beam for 30~s, the metastable fraction was reduced to less than 1\% during measurement. TSR EII results showed excellent agreement between theory and experiment for this ion. 

\subsection{B- to Ne-like ions $(2s^{2}\,2p^{q}, q=1-6)$}\label{subsec:bne}

\begin{figure}[t]
%\begin{center}
\begin{minipage}[t]{0.48\textwidth}
%\resizebox{\textwidth}{!}{\includegraphics{Fe91113_vs_cb.eps}}
\includegraphics[width=\textwidth, trim=8mm 0mm 3mm 2mm]{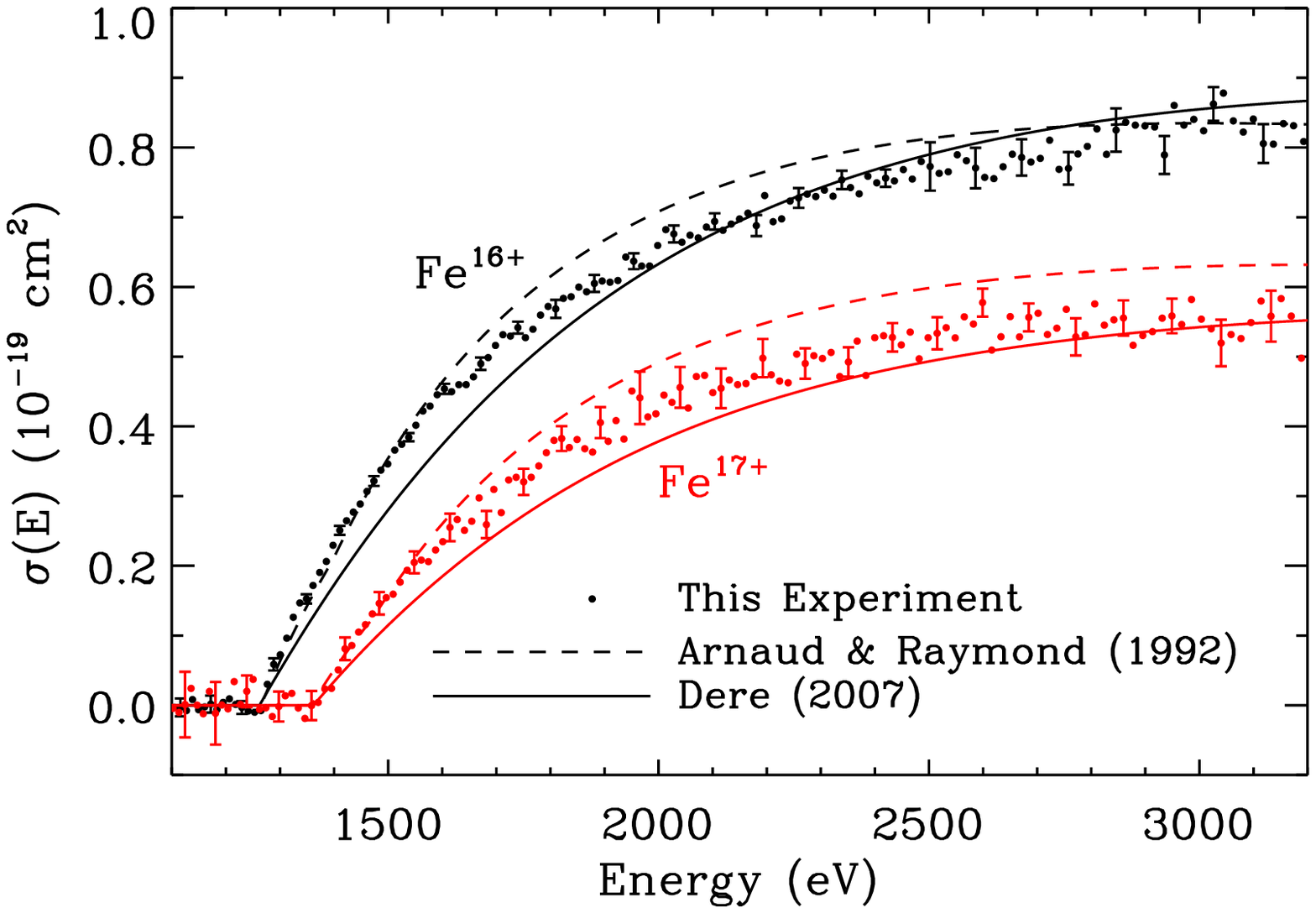}
\caption{\label{fig:fe1617} Fe$^{16+}$ and Fe$^{17+}$ EII cross sections from experiment ($\fullcircle$), Ref.~\cite{Arnaud:ApJ:1992} ($\dashed$), and Ref.~\cite{Dere:AA:2007} ($\full$). The error bars on selected points illustrate the 1$\sigma$ statsitical uncertainties. The systematic uncertainty for these EII measurements is about 15\%.}
\end{minipage}
\hfill
\begin{minipage}[t]{0.48\textwidth}
%\resizebox{\textwidth}{!}{\includegraphics{Fe91113_vs_ARDere.eps}}
\includegraphics[width=\textwidth, trim=8mm 0mm 3mm 2mm]{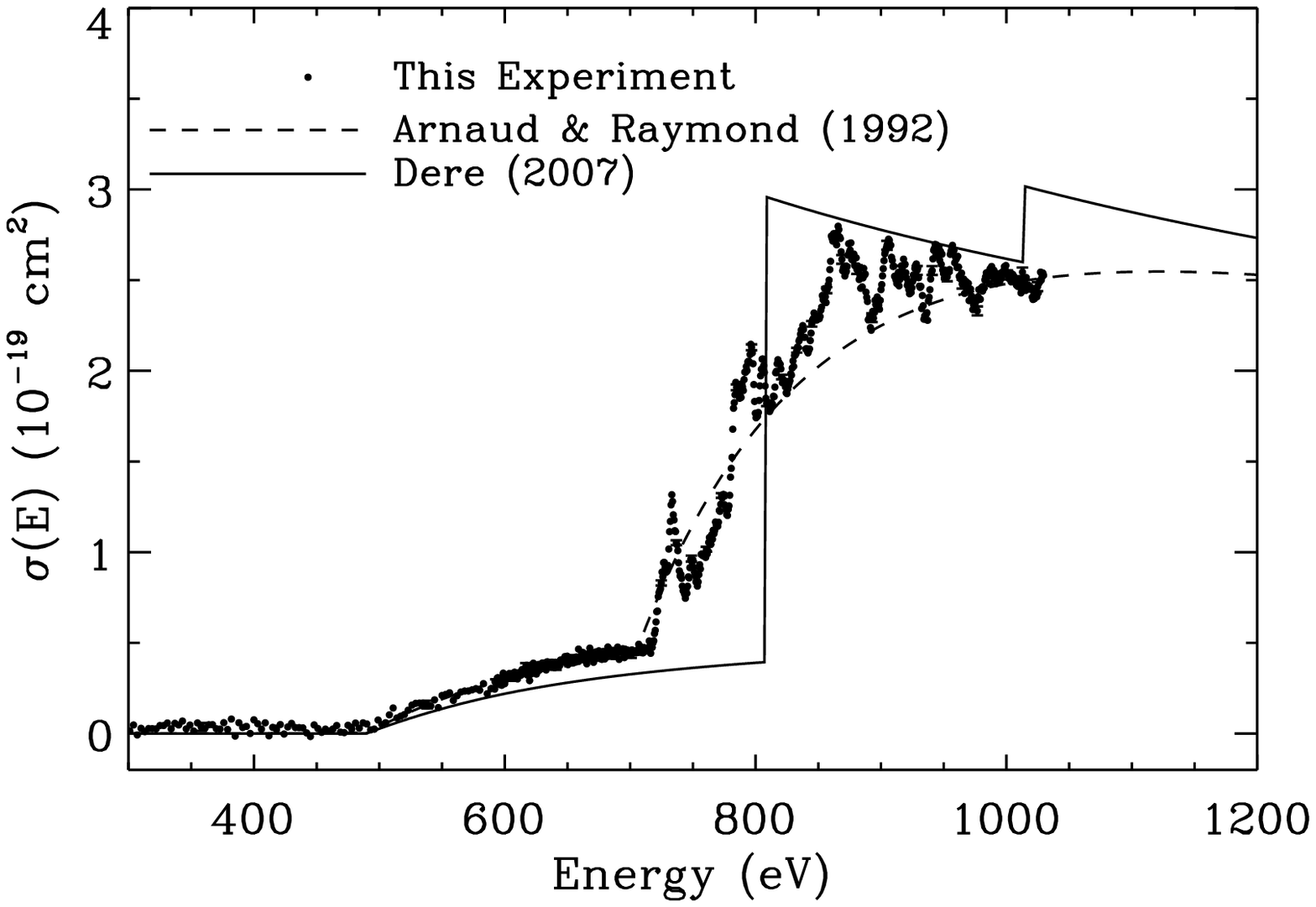}
\caption{\label{fig:fe15} Fe$^{15+}$ EII cross section from a TSR measurement ($\fullcircle$) \cite{Linkemann:fe15} compared to the theoretical cross sections of Arnaud \& Raymond ($\dashed$) \cite{Arnaud:ApJ:1992} and Dere ($\full$) \cite{Dere:AA:2007}. The error bars on the experimental data show the $1\sigma$ statistical uncertainties for selected points. The systematic uncertainty is $20\%$.}% The crossed beams cross sections are significantly larger than ours due to the presence of metastables in those experiments. }
\end{minipage}
\hfill
%\end{center}
\end{figure}

	For this group TSR measurements have been performed for B-like Mg$^{7+}$ \cite{Hahn:mg7}, F-like Fe$^{17+}$ \cite{Hahn:fe13}, and Ne-like Fe$^{16+}$ \cite{Hahn:fe13}, which all have a ground state configuration of $2s^2\,2p^{q}$ with $q=1,5,6$, respectively. Although C-like, N-like, or O-like ions were not measured, the results for B-like to F-like ions look qualitatively the same, and so it is expected that the cross sections for the unmeasured ions are similar. The Fe$^{17+}$ and Fe$^{16+}$ cross sections are illustrated in Figure~\ref{fig:fe1617}. 
	
	Overall, there is good agreement in magnitude between the measured cross section and theoretical prediction. One discrepancy is that the measured cross sections increase faster above the ionization threshold than predicted by Dere \cite{Dere:AA:2007}. However, the difference is at the level of the experimental systematic uncertainties and the measurement agrees with the older calculations \cite{Pindzola:NucFus:1987,Arnaud:ApJ:1992}. If the discrepancy is significant, it is likely due to excitation autoionization (EA) in which a $2s$ electron is excited to a level lying above the ionization threshold of the $2p$ electron. This process was neglected in the calculations, but measurements, discussed below, show that for $3s^2\,3p^q$ ions the analogous process is important. 

%\begin{figure}[th]
%\includegraphics[width=0.45\textwidth]{Fe1617.eps}%\hspace{2pc}%
%\begin{minipage}[b]{0.45\textwidth}\caption{\label{fig:fe1617} Fe$^{16+}$ and Fe$^{17+}$ ionization cross sections from %experiment ($\fullcircle$), Ref.~\cite{Arnaud:ApJ:1992} ($\dashed$), and Ref.~\cite{Dere:AA:2007} ($\full$)} %[b]before 14pc
%\end{minipage}
%\end{figure}

%\begin{figure}[t]
%\parbox{0.48\textwidth}{\includegraphics[width=0.48\textwidth]{Fe1617.eps}}
%\parbox{0.48\textwidth}{\caption{\label{fig:fe1617} Fe$^{16+}$ and Fe$^{17+}$ ionization cross sections from experiment ($\fullcircle$), Ref.~\cite{Arnaud:ApJ:1992} ($\dashed$), and Ref.~\cite{Dere:AA:2007} ($\full$)
%}}
%\end{figure}

\subsection{Na- and Mg-like ions $(3s^q, q=1,2)$}\label{subsec:namg}

	The measurement of Na-like Fe$^{15+}$ \cite{Linkemann:fe15, Linkemann:fe15a, Muller:IJMS:1999}, as well as Cl$^{6+}$ and Se$^{23+}$ \cite{Linkemann:thesis} were among the first EII measurements at TSR. More recently, Mg-like Fe$^{14+}$ has been measured \cite{Bernhardt:fe14}. The EII cross sections for these ions exhibit resonances, which are most prominent for Fe$^{15+}$ and other Na-like ions (Figure~\ref{fig:fe15}). Resonances are formed when dielectronic capture produces an excited state that decays by ejecting two electrons. The process can be either resonant excitation double autoionization (REDA) \cite{Lagattuta:PRA:1981,Muller:PRL:1988,Muller:AAMOP:2008} or resonant excitation auto-double ionization (READI) \cite{Henry:PRA:1982,Muller:PRL:1988a,Muller:AAMOP:2008}. For these ions, the dielectronic capture process occurs through the excitation of an $n=2$ electron and the resonances coincide with the various $n=2$ EA channels.
	
	The TSR measurements for Fe$^{15+}$ and Fe$^{14+}$ are in reasonable agreement with the theoretical cross sections of Dere \cite{Dere:AA:2007}. However, there is a significant discrepancy with the Arnaud \& Raymond data, which predicts a large contribution to the cross section from direct ionization of an $n=2$ electron \cite{Arnaud:ApJ:1992}. For Fe$^{14+}$, no corresponding increase was found experimentally. The same would likely be true for Fe$^{15+}$, but the measurements did not extend to high enough energy. The reason for the discrepancy is that the excited state formed by ionization of an $n=2$ electron autoionizes leading to a net double ionization. This process was accounted for in the Dere calculations, with which there is good agreement.

\subsection{Al- to Cl-like ions $(3s^{2}\,3p^{q}, q=1-5)$}\label{subsec:alcl}

\begin{figure}[t]
%\begin{center}
\begin{minipage}[t]{0.48\textwidth}
%\resizebox{\textwidth}{!}{\includegraphics{Fe91113_vs_cb.eps}}
\includegraphics[width=\textwidth, trim=8mm 0mm 3mm 2mm]{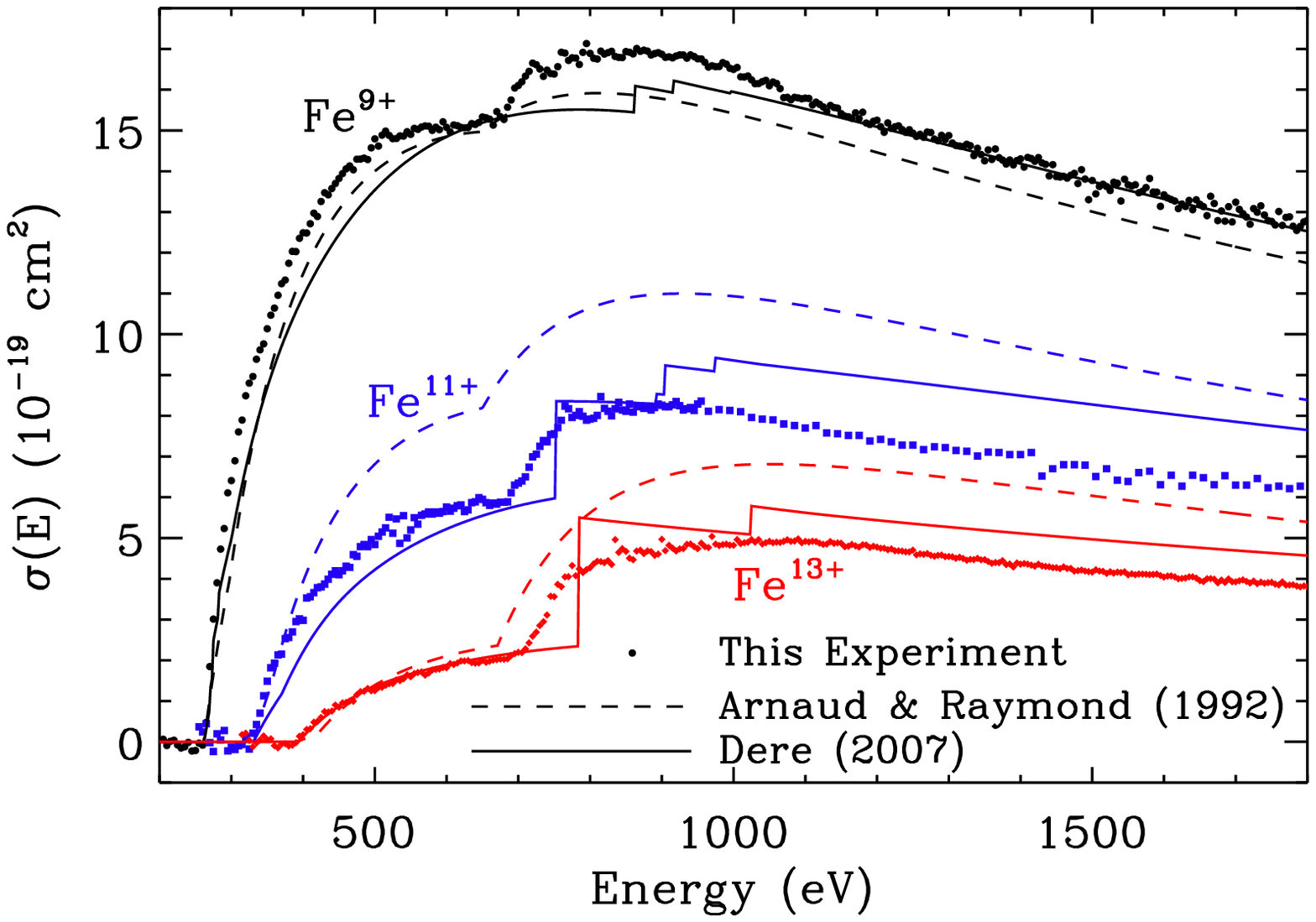}
\caption{\label{fig:fe91113} A comparison of TSR results for Fe$^{9+}$, Fe$^{11+}$, and Fe$^{13+}$ (filled symbols) to the recommended cross sections of Arnaud \& Raymond \cite{Arnaud:ApJ:1992} ($\dashed$) and Dere \cite{Dere:AA:2007} ($\full$). For the measurements the statistical uncertainties are smaller than the symbol size. The $1\sigma$ systematic uncertainties are 10 -- 15\%.}
\end{minipage}
\hfill
\begin{minipage}[t]{0.48\textwidth}
%\resizebox{\textwidth}{!}{\includegraphics{Fe91113_vs_ARDere.eps}}
\includegraphics[width=\textwidth, trim=8mm 0mm 3mm 2mm]{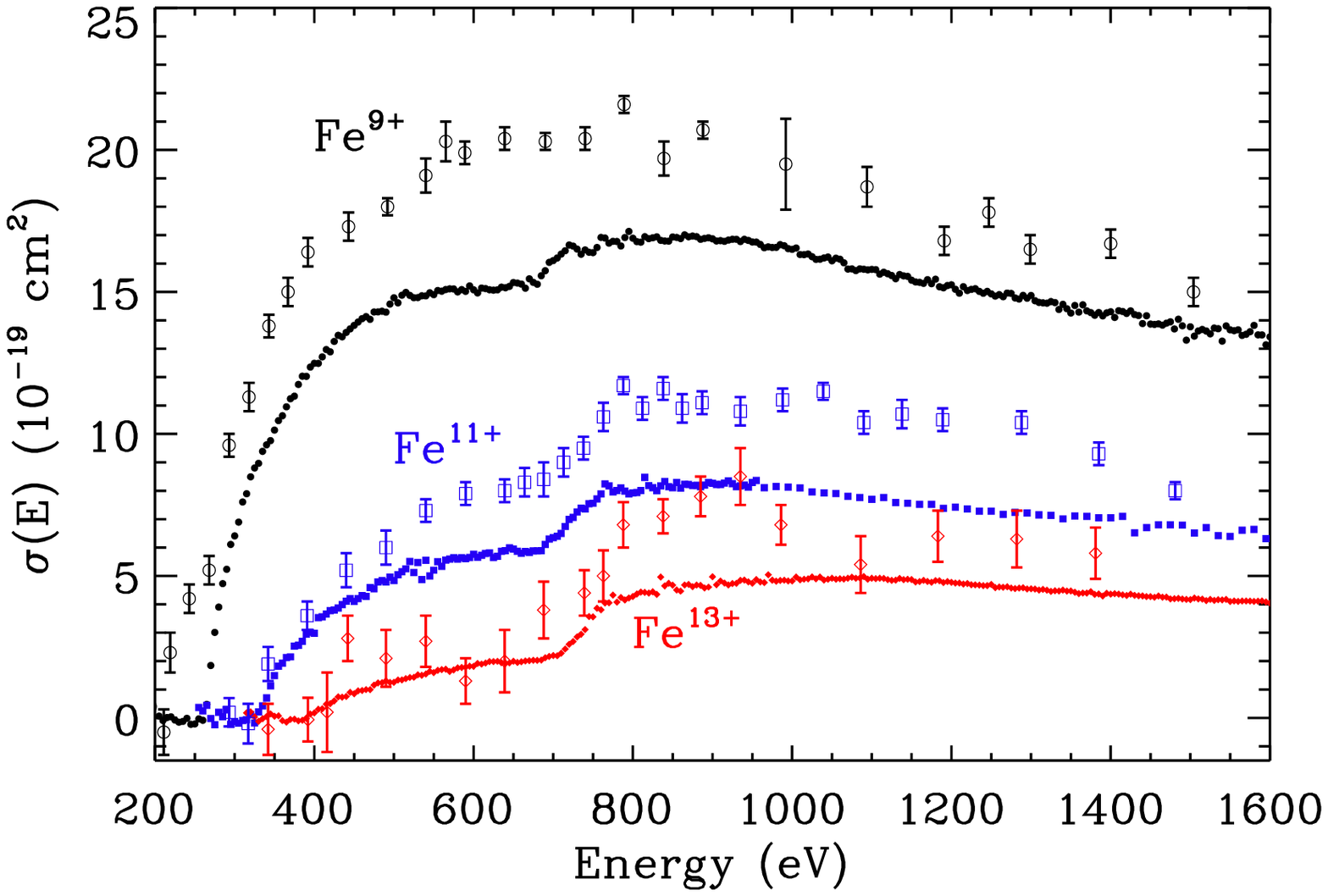}
\caption{\label{fig:tsrvscb} A comparison of the TSR results (filled symbols) to the crossed beams measurements (open symbols) of Gregory et al. \cite{Gregory:PRA:1986,Gregory:PRA:1987}, illustrating the effect of metastable ions. For the TSR data the statistical uncertainties are smaller than the symbol size.  The $1\sigma$ systematic uncertainties are 10 -- 15\%.}% The crossed beams cross sections are significantly larger than ours due to the presence of metastables in those experiments. }
\end{minipage}
\hfill
%\end{center}
\end{figure}

	EII was studied for Al-like Fe$^{13+}$ \cite{Hahn:fe13}, Si-like Fe$^{12+}$ \cite{Hahn:fe12}, P-like Fe$^{11+}$ \cite{Hahn:fe11}, S-like Fe$^{10+}$ \cite{Hahn:fe9}, and Cl-like Fe$^{9+}$ \cite{Hahn:fe9}. All these ions have a ground configuration of $3s^2\,3p^{q}$ for $q=1-5$, respectively. For these ions, several significant discrepancies with theory have been found (Figure~\ref{fig:fe91113}). 
	
	For all of these ions, except possibly Fe$^{13+}$, the experimental cross section rises faster near threshold than predicted by theory. We suggested that the reason for this was that the calculations neglect EA from excitation of the $3s$ electron to states that relax through autoionization \cite{Hahn:fe11}. Kwon \& Savin \cite{Kwon:PRA:2012} recently performed FAC calculations for Fe$^{11+}$ that included this EA channel and found good agreement with experiment near the ionization threshold. Fe$^{13+}$ may be an exception due to there being only one $3p$ electron or the effect may be masked by the 16\% systematic uncertainties for that measurement \cite{Hahn:fe13}. 
	
	Theory and experiment also showed differences in the EA from $n=2$ excitations. The cross section for each of these ions also has a significant contribution from $n=2\rightarrow3$ excitations occuring at about 700~eV. In some cases, such as for Fe$^{9+}$ and Fe$^{10+}$, this channel was neglected by theory \cite{Hahn:fe9}. At higher energies, theory predicts significant contributions from $n=2\rightarrow4$ and higher transitions. But EA from these transitions was found to be negligible. A probable reason for this discrepancy is that theory underestimates the branching ratio for radiative stabilization of the excited state. An alternate possibility is that the branching ratio for auto-double ionization is underestimated. However, TSR measurements for the double ionization cross sections do not find a corresponding increase from excitation double autoionization \cite{Hahn:fe9,Hahn:fe11,Hahn:fe12,Hahn:fe13}. Therefore, it seems more likely that the problem lies in the theoretical treatement of the radiative stabilization. 

	These data also demonstrate the value of the storage ring approach. Crossed beams experimental measurements exist for Fe$^{9+}$, Fe$^{11+}$, and Fe$^{13+}$. Those experimental cross sections are up to 35\% larger than the TSR measurements (Figure~\ref{fig:tsrvscb}). This is likely due to their having a large fraction of metastable ions in the ion beam, with the metastables having a larger cross section than the ground state ion. Consequently, CSD calculations that relied on the crossed beams EII data (e.g., \cite{Arnaud:ApJ:1992,Mazzotta:AAS:1998}) are innaccurate for these ions. 

%\clearpage
\subsection{K-like ions $(3s^2\,3p^6\,3d)$}\label{subsec:k}

	For K-like ions, TSR has measured EII of Fe$^{7+}$ \cite{Hahn:fe7}. Removing the metastables was a particular challenge for this ion. The ground state for Fe$^{7+}$ is $3s^2\,3p^6\,3d^{\;}\,^{2}D_{5/2}$. The longest-lived metastable levels are $3s^2\,3p^6\,3d^{\;}\,^{2}D_{3/2}$, which has a calculated lifetime of 11.0~s and $3s^2\,3p^5\,3d^2\,^{4}G_{11/2}$, with a calculated lifetime of 590~s \cite{Tayal:ApJ:2011}. The $^{2}D_{3/2}$ level population was greatly reduced by using a long, $>30$~s, storage time. Fortunately, this storage time also served to remove most of the $^{4}G_{11/2}$, despite the extremely long radiative lifetime of that level. This is probably because the $^{4}G_{11/2}$ metastables have a larger electron stripping cross section than the ground state ions. Thus, collisions with the residual gas attenuated the metastable beam faster than the ground state beam. The metastable fraction during measurement was estimated to be about $6\%$ with $\approx4\%$ from the $^{2}D_{3/2}$ level and $\approx2\%$ from the $^{4}G_{11/2}$ level.
	
%	 before measurement. Fortunately, this storage time also served to remove most of the $^{4}G_{11/2}$, despite the extremely long radiative lifetime of that level. The reason for this is probably that the $^{4}G_{11/2}$ metastables have a larger electron stripping cross section than the ground state ions. Thus, collisions with the residual gas attenuated the metastable beam faster than the ground state beam, resulting in a purer ground state ion beam. The metastable fraction during measurement was estimated to be about $6\%$ with about $\approx4\%$ from the $^{2}D_{3/2}$ level and about $\approx2\%$ from the $^{4}G_{11/2}$ level.
	
	Theoretical predictions overestimate the Fe$^{7+}$ cross section compared to TSR measurements (not shown here). The theory of Pindzola et al. \cite{Pindzola:NucFus:1987} used by Arnaud \& Raymond \cite{Arnaud:ApJ:1992} is 20\% to 40\% larger than experiment. The magnitude of the Dere \cite{Dere:AA:2007} cross section is within the systematic uncertainties of the experiment, but the energy dependence of the cross section is different. Part of the reason that Dere's result agrees in magnitude with the experiment is that it neglects direct ionization of a $3s$ electron, which was included in the other calculations \cite{Pindzola:NucFus:1987}. However, direct ionization of a $3s$ electron leads to an excited state that can relax radiatively or by autoionization. The contribution to single ionization versus double ionization depends on this unknown branching ratio. Another possible reason for the discrepancy between theory and experiment could be in the calculation of EA. For this ion, EA from $3p^{6}\,3d \rightarrow 3p^5\,3d\,nl$ transitions add greatly to the cross section near threshold. Theory may predict too large a contribution from these EA channels. 

%\begin{figure}[t]
%\parbox{0.48\textwidth}{\includegraphics[width=0.48\textwidth]{Fe7+_EII.eps}}
%\parbox{0.48\textwidth}{\caption{\label{fig:fe7} TSR measurement for Fe$^{7+}$ compared to the recommended data of Arnaud \& %Raymond~\cite{Arnaud:ApJ:1992} ($\dashed$), and Dere~\cite{Dere:AA:2007} ($\full$). 
%}}
%\end{figure}

\addtocounter{footnote}{1}
%\begin{savenotes}
\begin{table}[t]
\scriptsize
\caption{\label{table:ratecoeff} Polynomial fitting parameters $a_{i}$ to reproduce the scaled ionization rate coefficient $\rho$. The units of $\rho$ are $\mathrm{cm^{3}\,s^{-1}\,eV^{3/2}}$.}
\begin{center}
%\begin{minipage}{1.1\textwidth}
\begin{tabular}{llllllllllll}
\br
Sequence & Ion & Reference &  $E_{0}$~(eV) & $a_0$ & $a_1$ & $a_2$ & $a_3$ & $a_4$ & $a_5$ & $a_6$ \\
\mr
Be-like & S$^{12+}$ & \cite{Hahn:s12}  	& 652.2 		& 5.78526		& 6.83264		& -53.3550		& 148.287		& -176.114	& 75.6172 & \\
B-like & Mg$ ^{7+}$ & \cite{Hahn:mg7}		& 265.96		& 7.29335		& 10.2812		& -57.7081		& 125.183		& -117.002	& 36.6049 & \\
F-like & Fe$^{17+}$ & \cite{Hahn:fe13}	& 1357.8		& 17.4957		& 63.0820		& -485.718		& 1517.84		& -2127.02	& 1094.75	& \\
Ne-like & Fe$^{16+}$ & \cite{Hahn:fe13} & 1262.7		& 27.1411		& -24.4446	& 43.0958			& 74.8937		& -252.597	& 150.090		& \\
Na-like & Fe$^{15+}$ & \cite{Linkemann:fe15,Muller:IJMS:1999}	& 489.312		& -1.50046	& 65.8516		& -139.463		& 109.096 	& 31.7453		& -58.7546	& \\
Mg-like\footnotemark[1] & Fe$^{14+}$ & \cite{Bernhardt:fe14}	& 456.2	& 1.71700	& 39.4078		& 89.8572			& -517.427	& 716.450		& -328.129	& \\
Al-like & Fe$^{13+}$ & \cite{Hahn:fe13}	& 392.2			& 9.59988		& -53.2715	& 401.820 		& -1018.29	& 1115.84		& -455.515 & \\
Si-like & Fe$^{12+}$ & \cite{Hahn:fe12} & 361.04 		& 15.2688		& -84.5935	& 491.954			& -1156.69	& 1238.60		& -505.654 & \\
P-like 	& Fe$^{11+}$  & \cite{Hahn:fe11} & 330.79 		& 19.4438 	& -76.2632 	& 376.2632 		& -819.161 	& 827.141 	& -322.187 &\\
S-like 	& Fe$^{10+}$ & \cite{Hahn:fe9} 	& 290.25		& 18.3490 	& -20.2960 	& 16.0389 		& 122.420 	& -247.492 	& 124.888 & \\
Cl-like & Fe$^{9+}$ & \cite{Hahn:fe9}  	& 262.10 		& 28.4757 	& -76.3317 	& 280.844 		& -506.699 	& 465.127 	& -183.273 &  \\
K-like & Fe$^{7+}$ 	& \cite{Hahn:fe7}		& 151.060		& 43.4136 	& -329.950 	& 1987.92 		& -6312.25 	& 10734.7 	& -9190.33 & 3097.61 \\
\br
\end{tabular}
%\end{minpage}
\end{center}
\end{table}
\footnotetext[1]{Preliminary result.}
%\end{savenotes}

\section{Plasma rate coefficients}\label{sec:ratecoeff}

	In addition to benchmarking theory, the measured cross sections can be used directly in CSD calculations. Ionization rate coefficients for Maxwellian plasmas have been calculated based on the measured TSR EII cross sections. To facilitate their use in calculations, Table~\ref{table:ratecoeff} presents coefficients for a polynomial fit to the scaled rate coefficient $\rho(x)=10^{-6}\sum_{i}{a_i x^i}$.
%\begin{equation}	
%\rho(x)=10^{-6}\sum_{i}{a_i x^i}.
%\label{eq:polynomial}
%\end{equation}
The plasma rate coefficient as a function of electron temperature $\alpha_{\mathrm{I}}(T_{\mathrm{e}})$ is related to $\rho$ by \cite{Dere:AA:2007} 
\begin{equation}
\alpha_{\mathrm{I}}(T_{\mathrm{e}}) = t^{-1/2}E_0^{-3/2}E_{1}(1/t)\rho(x), 
\label{eq:invscalerate}
\end{equation}
where $E_{1}(1/t)$ is the first exponential integral and $t=k_{\mathrm{B}}T_{\mathrm{e}}/ E_0$ with $E_0$ being the ionization threshold and $k_{\mathrm{B}}$ the Boltzmann constant. The scaled temperature $x$ is given by
\begin{equation}
x = 1 - \frac{\ln 2}{\ln(t+2)}. 
\label{eq:invx}
\end{equation}
This expression can be inverted to obtain $T_{\mathrm{e}}$ from $x$ as
\begin{equation}
T_{\mathrm{e}} = \frac{E_0}{k_{\mathrm{B}}}\left[\exp\left(\frac{\ln 2}{1-x} \right) - 2 \right]. 
\label{eq:invscaletemp}
\end{equation}
The polynomial fit generally reproduces the experimental rate coefficient to better than 1\% accuracy for $T_{\mathrm{e}} \approx 10^{5}$ -- $10^{8}$~K. See the references for details about specific ions.

\section{Outlook}\label{sec:outlook}

	In the above discussion, we have skipped over Ar-like ions. An attempt was made to measure Ar-like Fe$^{8+}$, however, this ion has long-lived metastable levels that cannot be removed by storing the ions in TSR. The ground state for Fe$^{8+}$ is $3s^2\,3p^6\,^{1}S_{0}$. The first excited level is $3s^2\,3p^5\,3d^{\;}\,^{3}P_{0}$, which is forbidden to decay by a single photon transition and cannot be readily removed using an isotope because the hyperfine interaction is too weak. However, the $^{3}P_{0}$ level has a small statistical weight and so its population would be low. More of an issue is the $3s^2\,3p^5\,3d^{\;}\,^{3}F_{4}$ level, which has an estimated lifetime of $\sim 1000$~s \cite{Storey:AA:2002, Landi:ApJ:2013} and is expected to have an initial population similar to that of the ground state. From the experiment it was not possible to observe any difference in the cross section when using a 20~s versus 30~s initial storage time. This indicates that the metastables were not significantly removed over storage times of $\sim 30$~s. 
	
	Ions in isoelectronic sequences above K-like have also not been studied. These ions for astrophysically abundant elements generally have metastable levels with lifetimes that are too long to be removed with TSR storage times. For example Fe$^{6+}$, Fe$^{5+}$, and Fe$^{4+}$ all have metastable levels within the ground configuration with predicted lifetimes from 30~s to hundreds of seconds. Because of this, the existing crossed beams measurements are likely contaminated by metastables and should be verified with unambiguous storage ring measurements, but this cannot be done with TSR.
	
	The Cryogenic Storage Ring (CSR) currently under construction at the Max-Planck-Institut f\"ur Kernphysik will be able to overcome this limitation by having much longer storage times \cite{Krantz:JPCS:2011,vonHahn:NIMB:2011}. CSR will be cooled to $<10$~K and will have surfaces in the beamline at $< 2$~K, which will pump the chamber to extremely high vacuum through cryocondensation. The residual gas density will be $\sim 10^{3}$~$\mathrm{cm^{-3}}$, resulting in beam lifetimes of $\gtrsim 10^{3}$~s. Thus, CSR will allow us to extend storage ring EII measurements to ions with longer metastable lifetimes.
	
\section{Conclusion}\label{sec:sum}

	The storage ring approach provides valuable unambiguous EII data that cannot be obtained with other experimental geometries. These measurements provide needed benchmarks for theory and have identified several ways in which theoretical predictions can be improved. One improvement is that theory should not neglect EA from excitations within the same shell, such as $3s$ EA in $3s^2\,3p^q$ ions. Recent theoretical calculations accounting for this process have verified that including this channel provides more accurate theoretical data. Also, current theories predict too large of a cross section for EA from $n=2 \rightarrow n > 3$ transitions. This appears to be due to theory underestimating the branching ratio for radiative stabilization for such transitions. 
	
	Up to now, ions have been studied from almost every isoelectronic sequence from Li-like to K-like. In the future, we plan to continue this work in order to study EII for isoelectronic sequences higher than K-like. These ions are more complex and will provide new tests for EII theory.

\ack

	The author would like to thank all the collaborators who contributed to this work: Arno Becker, Dietrich Bernhardt, Manfred Grieser, Claude Krantz, Michael Lestinsky, Alfred M\"uller, Old\v{r}ich Novotn\'y, Roland Repnow, Stefan Schippers, Kaija Spruck, Andreas Wolf, and Daniel Wolf Savin. We appreciate also the efficient support by the MPIK accelerator and TSR groups for the beamtimes. This work has been supported in part by the NASA Astronomy and Physics Research and Analysis program, the NASA Solar Heliospheric Physics program, the Max Planck Society, and Deutsche Forschungsgemeinschaft. 
	
%\begin{center}
%\begin{table}[h]
%\caption{\label{opt}\cls\ class file options.}
%%\footnotesize\rm
%\centering
%\begin{tabular}{@{}*{7}{l}}
%\br
%Option&Description\\
%\mr
%\verb"a4paper"&Set the paper size and margins for A4 paper.\\
%\verb"letterpaper"&Set the paper size and margins for US letter paper.\\
%\br
%\end{tabular}
%\end{table}
%\end{center}

%\begin{figure}[h]
%\begin{minipage}{14pc}
%\includegraphics[width=14pc]{name.eps}
%\caption{\label{label}Figure caption for first of two sided figures.}
%\end{minipage}\hspace{2pc}%
%\begin{minipage}{14pc}
%\includegraphics[width=14pc]{name.eps}
%\caption{\label{label}Figure caption for second of two sided figures.}
%\end{minipage} 
%\end{figure}

%\begin{figure}[h]
%\includegraphics[width=14pc]{name.eps}\hspace{2pc}%
%\begin{minipage}[b]{14pc}\caption{\label{label}Figure caption for a narrow figure where the caption is put at the side of the %figure.}
%\end{minipage}
%\end{figure}

\section*{References}
\bibliography{Icpeac_EII}
\end{document}